\documentclass[
    ,final            
  ]
  {aipproc}

\layoutstyle{8x11double}

\begin{document}

\title{Neutron capture cross sections: \\ 
from theory to experiments and back}

\author{A.~Mengoni\footnote{\sf e-mail: alberto.mengoni@cern.ch}}
{address={CERN, CH-1211 Geneva 23, Switzerland, and \\
ENEA, Via Don Fiammelli, 2 - 40129 Bologna, Italy}
}

\begin{abstract}
The method for an experimental determination of the stellar enhancement factor
for the cross section of the $^{151}$Sm$(n,\gamma)$ reaction 
process is proposed. This study offered the
pretext for an excursus on the interconnections between
capture and dissociation reactions and the interplay between theory and
experiments in the determination of neutron capture cross sections.
\end{abstract}

\maketitle


\section{Introduction}
One of the most difficult issues for the
accurate determination of stellar reaction rates is the
evaluation of the so-called stellar enhancement factor (SEF). 
This factor simply relates the cross section measured
in the laboratory to the cross section in a stellar
environment, $\sigma^{*}$, by
$\sigma^{*} \equiv {\rm SEF} \times \sigma^{lab}$.
In capture reactions, this means that for the
determination of the SEF, an evaluation
of the capture cross section for nuclei in
excited states needs to be done. In same cases
the SEF is close to unity and the
effect on the stellar cross section is
marginal. However, there are several important
cases in which excited states of the target nuclei 
are strongly populated in a stellar environment. There,
an accurate determination of the SEF is mandatory.
Just to mention one example in which this is
the case one can consider the SEF of
the $\sigma_{n,\gamma}$ of the $^{186}$Os and
$^{187}$Os. The latter, in particular, has an
excited state at only 9.8 KeV excitation energy
and at stellar temperature of the order of
$kT \approx $ 30 KeV can be populated up to
50~\%. This implies a 20~\% difference
in the stellar Maxwellian averaged capture cross
section ratio of $^{186}$Os to $^{187}$Os. In turn,
if one uses the Re/Os clock for the age determination
of the galactic nucleosynthesis,
one obtains an increase of 2 Gyr in the age.
It is therefore necessary to obtain an accurate
determination of the SEF.
This is usually done in model calculations
based on the Hauser-Feshbach statistical model
theory (HFSM). Measurements of the capture cross
sections for target nuclei in their
ground-state allow for checking the accuracy of 
the model calculations. However, an experimental
determination of the SEF would strongly enhance
the reliability of the evaluated reaction rates 
in stellar environment.
The possibility to measure directly the
cross section from nuclei in excited states
is scarce if not completely null,
because nuclei in excited states (which are not
isomers) live a marginal fraction of seconds.
It is proposed here to use the inverse
reaction channel of the neutron capture
process to deduce the SEF for one important
neutron capture cross section: 
the $^{151}$Sm$(n,\gamma)$. This reaction cross section
has been measured recently at the CERN n\_TOF 
facility~\cite{Abbondanno(2004)}
as well as at the Forschungszentrum, Karlsruhe~\cite{Wisshak(2004)}. 
These measurements provide confidence on the experimental determination 
of the Maxwellian averaged cross section (MACS) 
for the ground-state capture.
\begin{figure}[b]
  \includegraphics[height=.15\textheight]{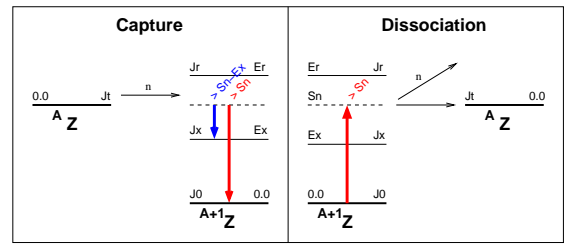}
  \caption{Schematic view of the neutron capture and
  dissociation processes}
\label{fig1}
\end{figure}
On the other hand, the SEF for this
reaction needs to be derived by theoretical
calculation which may produce different results. 
The possibility of an experimental determination of the SEF
seems therefore appealing.

\section{Direct and inverse reaction channels}
The time-reversal invariance of nuclear reactions 
which leads to the detailed-balance relation between
reactions in the "direct" and "inverse" channels has
been widely used for deriving reaction cross sections
which can be accessed in the laboratory only in one
of the time directions. A schematic view of the
capture reaction process and its time-reversals
invariant is shown in Figure~\ref{fig1}.
The relation between the cross sections
in the case of a neutron capture reaction process, 
$ n + ^{A}$Z $\rightarrow ^{A+1}$Z $+ \gamma$,
is simply given by
$$
\sigma_{n,\gamma} = \frac{k_{\gamma}^{2}}{k_{n}^2}
\frac{2J_{A+1}+1}{2J_{A}+1}
\sigma_{\gamma,n}
$$
where $k_{n}$ is the incident neutron wave number and 
$k_{\gamma} = \epsilon_{\gamma}/\hbar c$ is the
$\gamma$-ray wave number related to the $\gamma$-ray
transition energy $\epsilon_{\gamma}$.
The dissociation can be induced by real photons,
or by the virtual photon field generated by
a high-$Z$ target as in the Coulomb dissociation
process. The latter reaction process is nowadays
widely used in experiments with radioactive
ion beams, in which also the reaction
kinematics is inverted (see for example \cite{Fukuda(2004)}).
One important difference to notice between the two reaction
processes is that, while in the capture channel,
all the states in the $n + ^{A}$Z composite system
are populated, in the inverse dissociation process
only the ground-state of the composite system
is involved. Therefore, the time-reversal
invariance is fully symmetrical only
for ground-state transitions. It is also important
to note here that excited states in the
residual nucleus can be populated by the dissociation 
process. As will be shown in detail below, the proposal 
of the present work is based on this consideration.

On the experimental side, $(\gamma,n)$ reactions
have been used recently for the determination of
some capture reaction rates of unstable isotopes of
branching points in the $s$ process path, with
two different techniques.
An activation method based on a bremsstrahlung
photon spectrum such as the one of the S-DALINAC
at Darmstadt Technical University. In this case,
a photon beam with different bremsstrahlung
end-point energies is used to activate a
target material whose decay is detected off-line 
after the irradiation.
A second method is based on the use of
laser inverse-Compton scattering gamma-ray sources. 
In this case, a monochromatic gamma-ray beam is used
to induce prompt photo-neutron 
(or photo-$p$, photo-$\alpha$, etc.) emission in the
dissociation process. Both techniques are described
in a recent review paper~\cite{Utsunomiya(2004)} 
in which details of the methods
used to derive the capture reaction rate from
$(\gamma,n)$ measurements are given.
For both methods, a model calculation of the $(\gamma,n)$ 
cross section is required. The calculated cross section
is then normalized to the experimental data obtained
for the ground-state transitions and then 
the same normalization factor is used in the
evaluation of the cross section for transitions
to the other excited states.

\section{SEF: from theory to experiment}
The $^{151}$Sm$(n,\gamma)^{152}$Sm reaction has important implications
in nuclear astrophysics as it can be used to constrain the temperature
conditions of the He burning phase of AGB stars. With its
93 years $\beta$-decay half-life, $^{151}$Sm represents a branching point
in the $s$ process path which is influencing the isotopic abundance
ratio of the two $s$-only Gd isotopes $^{152}$Gd and $^{154}$Gd. Since
the $\beta$-decay half-life of $^{151}$Sm is temperature dependent, once that
its capture cross section is determined, a link between the temperature
conditions and the isotopic abundance ratio between the two Gd isotopes 
can be established.

The $^{151}$Sm$(n,\gamma)^{152}$Sm cross section has been recently
measured at the CERN n\_TOF facility~\cite{Abbondanno(2004)}. The
result for the MACS-30 (MACS at $kT = $ 30 keV) is $3100 \pm 160$ mb. 
This result has been confirmed in a measurement performed at 
FZK~\cite{Wisshak(2004)} in which the MACS-30 result obtained 
is $3031 \pm 68$ mb.
Theoretical estimates of the capture cross section based on HF statistical
model theory produces much lower values, typically around 2000 mb.
In any case, the energy dependence of the cross section is well reproduced
by the model calculation and, once that the MACS value is obtained
in the laboratory, the calculated cross section can be normalized.

\begin{figure}[ht]
  \includegraphics[height=.21\textheight]{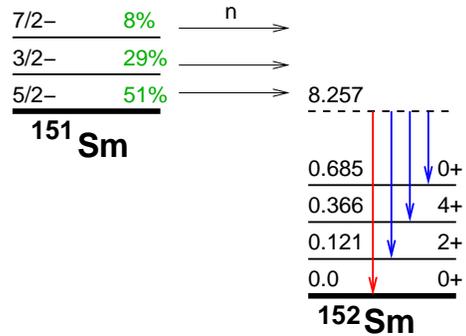}
  \caption{Schematic view of the neutron capture by $^{151}$Sm. 
  The population probabilities of the first three excitated target 
  states are indicated for a temperature of $kT = 30$ KeV.}
\label{fig2}
\end{figure}

$^{151}$Sm has several low-lying levels which can be
populated under stellar conditions. For example, the first excited state
at 4.8 keV is populated at a 30~\% level at temperatures of $kT = $ 30 KeV.
The determination of the SEF for the $^{151}$Sm$(n,\gamma)^{152}$Sm
reaction is therefore important for a reliable application
to AGB star modeling.
A schematic view of the situation in reported in Figure~\ref{fig2}.

As mentioned in the introduction, the SEF can be calculated. This
is normally done with the use of the Hauser-Feshbach statistical
model theory. The main ingredients in these calculations are the
optical model parameters representing the interaction of
the neutron with the target, furnishing the
neutron transmission coefficients for the neutron-nucleus
interaction. In addition, the representation of the density
of nuclear states at excitation energies up to the neutron
separation threshold and up to $\approx 1$ MeV above. Finally,
the electric dipole resonance parameters are used to evaluate
the $\gamma$-ray strength functions.

The NON-SMOKER calculation of the SEF for the
$^{151}$Sm$(n,\gamma)^{152}$Sm reaction is 0.87
at $kT = $ 30 KeV~\cite{Rauscher(2001)}, while in the calculation
performed for the present work we have obtained
0.93. This discrepancy is most likely due to
different parameterizations of the Hauser-Feshbach
theory which are used in model calculations.
Therefore, the accuracy of the SEF calculation
cannot guaranteed by the HFSM theory itself.

\section{The $^{152}$S\lowercase{m}$(\gamma,\lowercase{n})^{151}$S\lowercase{m} case}
\begin{figure}[ht]
  \includegraphics[height=.21\textheight]{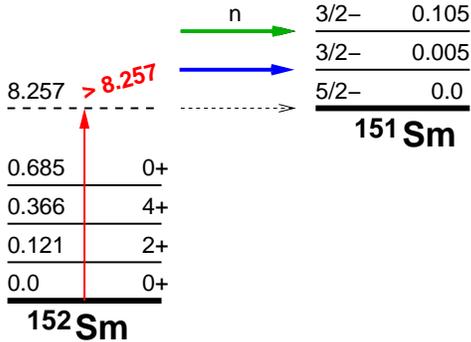}
  \caption{Dissociation of $^{152}$Sm into $^{151}$Sm$ + n$.}
\label{fig3}
\end{figure}

The $^{152}$Sm$(\gamma,n)^{151}$Sm reaction can be measured 
with both the techniques mentioned above.
There is, however, a peculiarity in this reaction.
In fact, starting from the $J^{\pi} = 0^{+}$ ground-state
of $^{152}$Sm, only excited states in $^{151}$Sm
can be populated by the strongest E1 transitions and
with $s$-wave neutrons in the continuum. In particular,
the $J^{\pi} = 5/2^{-}$ can only be populated with
$d$-wave neutrons left in the continuum. 
The $J^{\pi} = 3/2^{-}$ first excited state at $E_{x} = 4.82 $ KeV
in $^{151}$Sm is populated from the neutron emission threshold and
$s$-wave neutrons in the continuum. The next possible
excitation is the 5th excited state at $E_{x} = 104.8 $ KeV,
also a $J^{\pi} = 3/2^{-}$  state.
This situation is schematically shown in Figure~\ref{fig3}.

\begin{figure}[ht]
  \includegraphics[height=.23\textheight]{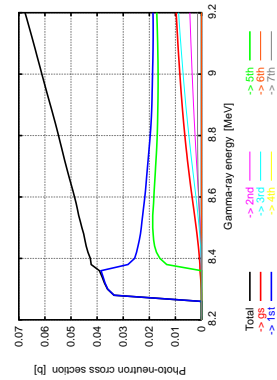}
  \caption{HFSM calculation of the $^{152}$Sm$(\gamma,n)^{151}$Sm
  cross section for dissociation into the different $^{151}$Sm$ + n$ channels.}
\label{fig4}
\end{figure}

The HFSM results of the calculation of the $(\gamma,n)$ reaction process
are shown in Figure~\ref{fig4}. There,
the cross sections $\sigma_{\gamma,n}$ for the various
reaction channels are shown in the energy region above
the neutron emission threshold. As expected, the
only open channel up to $\approx$ 100 KeV above
threshold is the dissociation leading to the 
$^{151}$Sm$(J^{\pi}=3/2^{-})$  state. Therefore, in an
actual experiment, this would produce the information
needed to derive the capture cross section of $^{151}$Sm
from its first excited state.

As the excitation energy increases, the second dissociation
channel at $E_{x} = $ 104.8 KeV opens up. Then, a measurement
of the $(\gamma,n)$ cross section above this energy range would
include the effect of this state and in turn, allow for an
experimental determination of the time-reversal neutron capture
process from this state. Only at higher excitation energies,
the effect of the ground-state starts to contribute, but only
to a few percent level.

The HFSM calculations have been performed using the nuclear level
density parametrization of reference \cite{Mengoni(1994)}. The
parameters have been fixed to reproduce the experimental level
spacings at the neutron separation energies which, in the
present case, are available for both the isotopes 
involved~\cite{Mughabghab(1981)}.
The Moldauer~\cite{Moldauer(1963)} optical model parameters have 
been used for the neutron transmission function calculations 
and experimental parameters of the Giant Dipole 
Resonance~\cite{Carlos(1974)} for the calculation of 
the $\gamma$-ray strength function. The cross section has been
renormalized in order to reproduce the measured 
$^{151}$Sm$(n,\gamma)^{152}$Sm MACS-30. This normalization factor
turned out to be 1.7 for the present HFSM calculations as well
as for the NON-SMOKER result~\cite{Rauscher(2001)}.

The absolute value of the dissociation cross section is of the same
order as that of measurements already performed (see for example the report
on the $^{186}$W$(\gamma,n)$ measurement~\cite{Sonnabend(2003)}),
making the proposal for an actual experiment quite realistic.

\section{Conclusion}

The experimental determination of the SEF for the
$^{151}$Sm$(n,\gamma)$ could be possible by a measurement
of the time-reversal invariant reaction process, the 
$^{152}$Sm$(\gamma,n)^{151}$Sm reaction. This reaction
can only populate excited states in the $^{151}$Sm$ + n$
residual channel for excitation energies just
above threshold. The technique has been used for the
determination of capture cross sections of unstable
nuclei at $s$ process branching points and could, in
this specific case lead to the experimental determination
of a SEF for the first time.


\begin{theacknowledgments}
I would like to thank Franz K\"{a}ppeler for several
fruitful discussions on the subject of the present work. 
\end{theacknowledgments}


\bibliographystyle{aipprocl} 


\hyphenation{Post-Script Sprin-ger}

\end{document}